\documentclass{aa}  

\usepackage[varg]{txfonts}

\usepackage{natbib}
\usepackage{graphicx}
\usepackage{dcolumn}
\usepackage{bm}
\usepackage{amssymb,amsmath,bm}
\usepackage{color}
\usepackage[colorlinks,linkcolor=red,citecolor=blue,urlcolor=blue ]{hyperref}
\usepackage{multirow}
\usepackage[utf8]{inputenc}
\usepackage{enumitem}

\graphicspath{%
	{./figures/} %
}	

\begin{document}

\defcitealias{Planck2018}{Planck~Collaboration~2020}
\defcitealias{Planck16}{Planck~Collaboration~2016}
\defcitealias{Planck13:cosmology}{Planck~Collaboration~2014}
\defcitealias{eBOSS20}{eBOSS~Collaboration~2020}

\title{Imprint of baryons and massive neutrinos on velocity statistics}
\titlerunning{Imprint of baryons and massive neutrinos on velocity statistics}

\author{Joseph Kuruvilla\inst{1}, Nabila Aghanim\inst{1} \and Ian G. McCarthy\inst{2}}

\offprints{Joseph Kuruvilla, \email{joseph.kuruvilla@universite-paris-saclay.fr}}

\institute{Universit\'e Paris-Saclay, CNRS,  Institut d'Astrophysique Spatiale, B\^atiment 121, 91405, Orsay, France
\and
Astrophysics Research Institute, Liverpool John Moores University, 146 Brownlow Hill, Liverpool L3 5RFF, UK}

\date{Received 2020; accepted}
\abstract{
\noindent
We explore the impact of baryonic effects (namely stellar and AGN feedback) on the moments of pairwise velocity using the Illustris-TNG, EAGLE, cosmo-OWLS, and BAHAMAS suites of cosmological hydrodynamical simulations. The assumption that the mean pairwise velocity of the gas component follows that of the dark matter is studied here at small separations, and we find that even at pair separations of 10-20 $h^{-1}\mathrm{Mpc}$ there is a 4-5\% velocity bias. At smaller separations, it gets larger with strength varying depending on the subgrid prescription. By isolating different physical processes, our findings suggest that the large scale velocity bias is mainly driven by stellar rather than AGN feedback. If unaccounted for, this velocity offset could possibly bias  cosmological constraints from the kinetic Sunyaev-Zel'dovich effect in future cosmic microwave background (CMB) surveys. Furthermore, we examine how the first and the second moment of the pairwise velocity are affected by both the baryonic and the neutrino free-streaming effects for both the matter and gas components. For both moments, we were able to disentangle the effects of baryonic processes from those of massive neutrinos; and below pair separations of 20 $h^{-1}\mathrm{Mpc}$, we find that these moments of the pairwise velocity decrease with increasing neutrino mass. Our work thus paves a way in which the pairwise velocity statistics can be utilised to constrain the summed mass of neutrinos from future CMB surveys and peculiar velocity surveys.}

\keywords{cosmology: large-scale structure of the Universe -- cosmology: theory}

\maketitle
\section{Introduction}

Over the last decade or so, cosmology has evolved to a state where we are able to precisely constrain the cosmological parameters with the help of galaxy redshift surveys \citepalias[e.g.][]{eBOSS20}, gravitational lensing surveys \citep[e.g.][]{Kids1000} and cosmic microwave background (CMB) experiments \citepalias[e.g.][]{Planck2018}. Some of the outstanding questions which remain are regarding the dark sector, including determining the nature of dark energy and the summed mass of neutrinos. In order to answer these questions, peculiar velocity surveys provide a complementary avenue to further our understanding.  Forthcoming peculiar velocity surveys, such as the Taipan galaxy survey\footnote{\url{https://www.taipan-survey.org}} \citep{Taipan}, the Widefield ASKAP  L-band  Legacy  All-sky  Blind Survey\footnote{\url{https://www.atnf.csiro.au/research/WALLABY/}} \cite[WALLABY,][]{Wallaby-20}, and the Westerbork Northern Sky HI Survey (WNSHS), promise to be competitive as cosmological probes for very low redshifts with the respect to current galaxy clustering surveys \citep{Koda+14,Howlett+17}.   

The current lower limit of sum of neutrino masses, $M_\nu = \sum m_\nu \gtrsim 0.06$ eV, comes from the neutrino oscillation experiments \citep[e.g.][]{Forero+14,Gonzalez-Garcia+16,Capozzi+17,Salas+17}. Massive neutrinos, unlike dark matter or baryons, have large thermal velocities which imprint distinct signatures on the cosmological observables. Leveraging this by combining different cosmological probes serves as an upper limit on the  neutrino mass constraints.  Depending on which datasets are combined and how the analysis is done, the current upper limit on the summed neutrino mass ranges from 0.12 eV up to $\approx$0.4 eV (e.g~\citealt{diValentino+16,Vagnozzi+17,McCarthy+18,Poulin+18,Palanque+20,Ivanov+20};~\citetalias{Planck2018}).  The impact of massive neutrinos on clustering statistics in real and redshift space has been studied \citep[e.g.][]{Saito+08,Wong08,Castorina+15,Navarro+18,Garcia+19}. Forthcoming redshift surveys will provide tighter constrains on $M_\nu$ using two-point and three-point galaxy clustering statistics \citep{ChudaykinIvanov19}. Furthermore, the bispectrum should help breaking neutrino mass and $\sigma_8$ degeneracy \citep{Hahn+20}.
In addition to clustering statistics, the one-point probability distribution function of the total matter has been shown sensitive to neutrino mass and could provide strong constraints \citep{Uhlemann+19}. 

In this era of precision cosmology, it is important to consider the effects of baryons and processes associated with galaxy formation (e.g. cooling and feedback) on cosmological observables, particularly as we push the analyses to smaller, `non-linear' scales.  It has been shown, for example, that dark energy constraints can be biased by baryonic effects if they are unaccounted for (e.g. \citealt{Semboloni+11,Copeland+18}). So far, much of the attention has been focused on the impact of baryons on the clustering statistics, for example in the case of two-point statistics in Fourier space  \citep[e.g.][]{vanDaalen+11,Chisari+18,Chisari+19,Schneider+19,vanDaalen+20} and in configuration space \citep{vanDaalen+14}.  \cite{Mummery+17} has shown that the effects of baryon physics (feedback) and neutrino free-streaming are separable (multiplicative), to typically a few percent accuracy, in their effects on the clustering statistics, even on deeply non-linear scales.  The effect of baryons on the matter bispectrum have also recently been examined \citep[e.g.][]{Foreman+19}.  These studies were done with the aid of cosmological hydrodynamical simulations.  Note that an alternative approach is to use the so-called ``baryonic correction model'', where the baryonic effects are parameterised based on physically-motivated parameters and used to modify the outputs of cosmological $N$-body simulations \citep[e.g.][]{SchneiderTeyssier15,Schneider+19,Arico+19,Arico+20}. 

The main aim of the present paper is to understand the effects of baryonic processes and massive neutrino effect on velocity statistics, namely on the first two moments of the pairwise velocity statistics, at pair separations below 20 $h^{-1}\mathrm{Mpc}$.  Relatively little attention has been devoted to the impact of baryons and neutrinos on the velocity statistics to date, particularly in comparison to the numerous studies on the spatial distribution of matter/haloes.  As we will describe in the following paragraphs, the pairwise velocity has applications mainly in three areas of cosmology: 

(i) Galaxy clustering: the observed positions of galaxies are perturbed from their true positions due to their peculiar velocities, an effect which is known as `redshift-space distortions' (RSD). These distortions can be leveraged to accurately constrain the growth rate of structure, and hence cosmological parameters, by measuring correlation functions in redshift space \citep{PercivalWhite09}.  In configuration space clustering, the state-of-the-art modelling is based on the `streaming model' \citep{Peebles80, Fisher95, Scoccimarro04, KuruvillaPorciani18, VlahWhite18}, recently generalised to $n$-point correlation function in redshift space \citep{KuruvillaPorciani20}. In two-point clustering, it provides a framework to map the two-point correlation function in redshift space, which is obtained as the integral of the real-space isotropic correlation function with the line-of-sight (los) pairwise velocity distribution. The key element in this streaming model framework is the pairwise los velocity distribution. Thus, understanding how the 
pairwise velocity statistics are affected by baryons and neutrinos will further help in modelling small-scale redshift-space clustering statistics. Within the streaming model framework, \cite{AvilesBanerjee20} have recently studied the effects of neutrinos on pairwise velocity statistics and redshift-space correlation function using Lagrangian perturbation theory, above scales of 20 $h^{-1} \mathrm{Mpc}$.

(ii) Peculiar velocity surveys: direct measurements of the peculiar velocity can be achieved through redshifts and distances determined through scaling relations, such as the Tully-Fisher \citep{TullyFisher77} or the Fundamental Plane relations \citep{DjorgovskiDavis87,Dressler+87}.  These direct peculiar velocity surveys are shallow and thus offers an opportunity to probe the peculiar velocities in the nearby Universe.  In \cite{Dupuy+19}, the mean pairwise velocity estimator was used to constrain the growth rate of the structure using the Cosmicflows-3 dataset \citep{Cosmicflows3}.

(iii) Kinetic Sunyaev-Zeldovich (kSZ) effect: a secondary anisotropy where CMB photons are scattered off free electrons which are in motion.  This results in a Doppler shift, thus preserving the blackbody spectrum of the CMB \citep{SZ72, SZ80}. The fluctuation in the CMB temperature can be written as
\begin{equation}
    \frac{\Delta T_{\mathrm{kSZ}}}{T_{\mathrm{cmb}}} = \sigma_{\mathrm{T}}\int\mathrm{d}l \ n_e \left(\frac{\boldsymbol{v}_{\mathrm{e}}\cdot \hat{\boldsymbol{n}}}{c}\right) \simeq - \tau \left(\frac{\boldsymbol{v}_{\mathrm{e}}\cdot \hat{\boldsymbol{n}}}{c}\right) \, , 
\end{equation}
where $\sigma_{\mathrm{T}}$ is the Thomson scattering cross section, $n_e$ is the electron number density, $\boldsymbol{v}_{\mathrm{e}}$ is the velocity of the free electrons, $\hat{n}$ is the unit vector along the line-of-sight, and $\tau=\sigma_{\mathrm{T}}\int \mathrm{d}l\ n_e$ is the optical depth.
It is also one of the techniques through which we can measure the peculiar velocities of objects at cosmological distances. However, the signal from kSZ effect is very weak, hence detections for individual objects have proven to be difficult so far.  Currently, detections of kSZ are mainly limited to the mean pairwise velocity, as it can be measuring through stacking techniques to boost the signal.  The first detection of kSZ effect through the pairwise mean velocity was by \cite{hand+12} using the pairwise velocity estimator developed by \cite{Ferreira+99}. Further evidence for kSZ using pairwise velocities were presented in \citetalias{Planck16};~\cite{Soergel+16,Bernardis+17,Li+17}.  It has been shown that the mean radial pairwise velocity measured from the kSZ effect is capable of constraining alternative theories of gravity and dark energy \citep{BhattacharyaKosowsky07,BhattacharyaKosowsky08,KosowskyBhattacharya09,Mueller+15a}, in addition placing constraints on the summed mass of neutrinos \citep{Mueller+15b}.  Alternatively, kSZ effect have been detected by correlating CMB maps with reconstructed velocity field \cite[e.g.][]{Schaan+16,Tanimura+20,Nguygen+20}.  Future CMB surveys, such as the the Simons Observatory\footnote{\url{https://simonsobservatory.org/}} \cite[SO,][]{SO}, CMB-S4\footnote{\url{https://cmb-s4.org/}} \citep{CMBS4} and CMB-HD \citep{CMBHD-I, CMBHD-II}, will be able to measure the kSZ effect, and in turn the pairwise velocity statistics, much more precisely.

As already noted, the aim of this paper is to disentangle the effects baryonic processes and massive neutrino on the first two moments of the pairwise velocity.  We also examine the typical assumption that the pairwise velocity of gas follows that of the dark matter for the mean pairwise velocity.  This assumption been tested for pairwise kSZ signal in \cite{Flender+16} using halos from $N$-body simulations and adding a gas profile following a model introduced in \cite{Shaw+10}. However in this paper we follow the gas particles directly from hydrodynamical simulations.

The paper is structured as follows. In Sect.~\ref{sec:sims}, we briefly summarise the various hydrodynamical simulations employed in this work. In Sect.~\ref{sec:radial-pairwise}, we introduce the radial pairwise velocity. We  introduce its first moment, the mean radial pairwise velocity, and how it is impacted by different baryonic processes in Sect.~\ref{sec:mean-radial}. We focus on how massive neutrinos affect the first moment in Sect.~\ref{sec:neutrino-mean}, and the second moment in Sect.~\ref{sec:dispersion-radial}. Finally, we summarise our findings in Sect.~\ref{sec:conclusion}.

\section{Simulations}
\label{sec:sims}
In this work, we make use of four suites of hydrodynamical simulations to measure the pairwise velocity statistics: Illustris-TNG, EAGLE, cosmo-OWLS and BAHAMAS.  We briefly describe these simulations below.
\begin{table*}[]
    \caption{Characterisation of the various simulations used in this work. BAHAMAS (0) refers to the reference simulation with zero neutrino mass. The length of the simulation is denoted by $L_{\mathrm{box}}$. While $m_{\mathrm{DM}}$ and $m_{\mathrm{b}}$ denotes the mass of the dark matter and baryon species respectively.}
    \label{tab:sim_properties}
    \centering
    \renewcommand{\arraystretch}{1.3}
    \begin{tabular}{lccccc}
    \hline
    \hline
    \multicolumn{1}{c}{Simulation} & \multicolumn{1}{c}{Hydrodynamical code} & \multicolumn{1}{c}{$L_\mathrm{box}$} & \multicolumn{1}{c}{$m_{\mathrm{DM}}$} & \multicolumn{1}{c}{$m_{\mathrm{b}}$} & \multicolumn{1}{c}{Cosmology} \\
     &  & \multicolumn{1}{c}{$[h^{-1}\mathrm{Mpc}]$} & \multicolumn{1}{c}{$[\ h^{-1}\mathrm{M}_\odot]$} & \multicolumn{1}{c}{$[\ h^{-1}\mathrm{M}_\odot]$} & \\
     \hline
    EAGLE & \textsc{Gadget} & 67.8 & $6.7\times10^6$ & $1.8\times10^6$ & \textit{Planck} 2013 \\ 
    Illustris-TNG100 & \textsc{arepo} & 75.0  & $5.1\times10^6$ & $0.9\times10^6$ & \textit{Planck} 2016 \\
    Illustris-TNG300 & \textsc{arepo} & 205.0  & $3.9\times10^7$ & $7.4\times10^6$ & \textit{Planck} 2016 \\
    cosmo-OWLS & \textsc{Gadget} & 400.0 & $3.7\times10^9$  & $7.5\times10^8$  & \textit{WMAP}7\\    
    BAHAMAS (0) & \textsc{Gadget} & 400.0 & $3.8\times10^9$ & $7.6\times10^8$ & \textit{WMAP}9\\
    \hline
    \\
    \end{tabular}
    \\
\end{table*}

\begin{enumerate}
\item Illustris-TNG: `The Next Generation Illustris Simulations' \citep{TNG-I,TNG-II,TNG-III,TNG-IV,TNG-V} is a suite of cosmological simulations run using the moving mesh code \textsc{arepo}.  It is a successor to the Illustris simulation \citep{Vogelsberger+14a,Vogelsberger+14b,Genel+14,Sijacki+15}. The subgrid physics has been updated from the original Illustris with changes in AGN feedback, galactic winds and inclusion of magnetic fields, which are described in detail in \cite{Weinberger+17} and \cite{Pillepich+18}. The feedback processes were calibrated to roughly reproduce several observed properties, such as the galaxy stellar mass function and the stellar-to-halo mass relation (see \citealt{Pillepich+18} for details).  This suite has simulations with three different volumes $50^3, 100^3$ and $300^3\ \mathrm{Mpc}^3$. In this work, we make use of the simulation boxes with side lengths of $100$ and $300\ \mathrm{Mpc}$, which have $1820^3$ and $2500^3$ tracer (dark matter [DM] and gas) particles, respectively. The simulation suite uses \textit{Planck} 2016 cosmological parameters, namely $\{\Omega_{\mathrm{m}},\Omega_{\mathrm{b}},\Omega_{\Lambda},h,n_{\mathrm{s}},\sigma_8\}=\{0.3089, 0.0486, 0.6911,0.6774,0.9667,0.8159\}$.

\item cosmo-OWLS \citep{LeBrun+14,McCarthy+14} is is an extension to the OverWhelmingly Large Simulations (OWLS) project \citep{Schaye+10}. Unlike the OWLS runs, most of which had boxes of length 100 Mpc $h^{-1}$, the cosmo-OWLS runs have boxes of length 400 Mpc $h^{-1}$. This suite of simulations were run using the smoothed  particle  hydrodynamics (SPH) code \textsc{gadget}-3 \citep{Springel+01,Springel05}. The details about the subgrid physics are given in \cite{LeBrun+14}, and we make use of five variants of the subgrid physics as listed in their Table 1. Unlike the other suites used here, no attempt was made to calibrate the feedback to match particular observations with cosmo-OWLS.  It was aimed at exploring the impact of large variations in the subgrid physics, including turning on or off physics such as radiative cooling and AGN feedback.
The simulation suite adopts a \textit{WMAP}7 cosmology, which is given by $\{\Omega_{\mathrm{m}},\Omega_{\mathrm{b}},\Omega_{\Lambda},h,n_{\mathrm{s}},\sigma_8\}=\{0.2720, 0.0455, 0.7280, 0.7040, 0.9670,0.8100\}$.

\item Evolution and Assembly of GaLaxies and their Environments \citep[EAGLE,][]{Schaye+15,Crain+15} is a set of cosmological hydrodynamical simulations evolved using \textsc{gadget}-3. The implemented subgrid physics is descended from OWLS but with several improvements as detailed in \cite{Schaye+15}.  The stellar and AGN feedback was calibrated to reproduce the present-day galaxy stellar mass function and the size--mass relation of galaxies.  The hydro solver scheme was also modified from classic SPH to the pressure--entropy `Anarchy' scheme, also described in the above references.
In this work, we make use of the box with volume $67.77^3\ h^{-3}\mathrm{Mpc}^3$ and with $2\times1504^3$ particles. The simulation suite adopts a \textit{Planck} 2013 \citepalias{Planck13:cosmology} cosmology, which is given by $\{\Omega_{\mathrm{m}},\Omega_{\mathrm{b}},\Omega_{\Lambda},h,n_{\mathrm{s}},\sigma_8\}=\{0.3070, 0.04825, 0.6930,0.6777,0.9611,0.8288\}$.

\item BAHAMAS is a suite of cosmological hydrodynamical simulation with a volume of $400^3\ h^{-3}\mathrm{Mpc}^3$ \citep{McCarthy+17,McCarthy+18}. This was also run using \textsc{gadget}-3. It follows the evolution of $1024^3$ DM and gas particles. Hence the mass resolution is lower than EAGLE or Illustris-TNG but is approximately the same as cosmo-OWLS.  And like cosmo-OWLS, it follows significantly larger volumes than EAGLE or Illustris.  The subgrid physics is based on the OWLS and cosmo-OWLS projects.  However, unlike OWLS and cosmo-OWLS, the feedback was explicitly calibrated to reproduce the observed present-day galaxy stellar mass function and the amplitude of the hot gas-halo mass relation of groups and clusters.  As BAHAMAS has the most realistic representation of baryons on large scales (including the gas fractions of massive groups and clusters), we expect the impact on large-scale structure to be more realistic for BAHAMAS.  The reference simulation we use adopts the \textit{WMAP}9 cosmology, which is given by $\{\Omega_{\mathrm{m}},\Omega_{\mathrm{b}},\Omega_{\Lambda},h,n_{\mathrm{s}},\sigma_8\}=\{0.2793,0.0463, 0.7207, 0.7000, 0.9720, 0.8211\}$ with massless neutrinos.

We also use an extension of BAHAMAS that includes massive neutrinos (see \citealt{McCarthy+18} for details). It consists of four simulations ranging from the lowest summed neutrino masses ($M_{\nu}$) of 0.06 eV up to 0.48 eV in factors of 2. The massive neutrinos were implemented keeping all the cosmological parameters fixed apart from $\sigma_8$ (note that $A_s$, the amplitude of the primordial power spectrum was kept fixed at the CMB value, consequently the inclusion of massive neutrinos lowers the $\sigma_8$) and the cold matter density ($\Omega_{\mathrm{cdm}}$), which was was decreased to ensure that the Universe is flat, where $\Omega_{\Lambda} + \Omega_{\mathrm{m}} = 1$ and $\Omega_{\mathrm{m}} = \Omega_{\mathrm{cdm}} + \Omega_{\nu} + \Omega_{\mathrm{b}}$. The neutrino density ($\Omega_{\nu}$) is related to $M_{\nu}$ by the relation $\Omega_{\nu} = M_{\nu}/(93.14\ h^2 \mathrm{eV})$. Thus, the BAHAMAS simulation explores $\Omega_\nu$ for a range of 0.0013 to 0.0105. This suite allows us to study the degeneracy between baryonic physics and massive neutrino effects on the pairwise velocity statistics in a systematic way.

\end{enumerate}

As noted above, the different sets of simulations have been calibrated employing various strategies. The characterisation of the simulations in terms of the box size, number of particles and mass resolution also varies between suites, as shown in Table~\ref{tab:sim_properties}. Despite the different underlying cosmological models in these simulations, we neglect the impact of cosmology on the baryonic effects.  Previous work based on extensions of the BAHAMAS suite (e.g., \citealt{Mummery+17,Stafford+20,Pfeifer+20}) have shown that the effects of baryon physics are separable from changes in cosmology at the few percent level for most statistics.  We have also verified (below) that the impact of fixed baryon physics on the pairwise velocity statistics is unaffected as the cosmology is changed to increase the summed mass of neutrinos.

It should be noted that there are corresponding collisionless simulations for each of the hydrodynamical runs above, including all of the massive neutrino cases.

\section{Radial pairwise velocity}
\label{sec:radial-pairwise}
\begin{figure}
    \includegraphics[width=0.99\columnwidth]{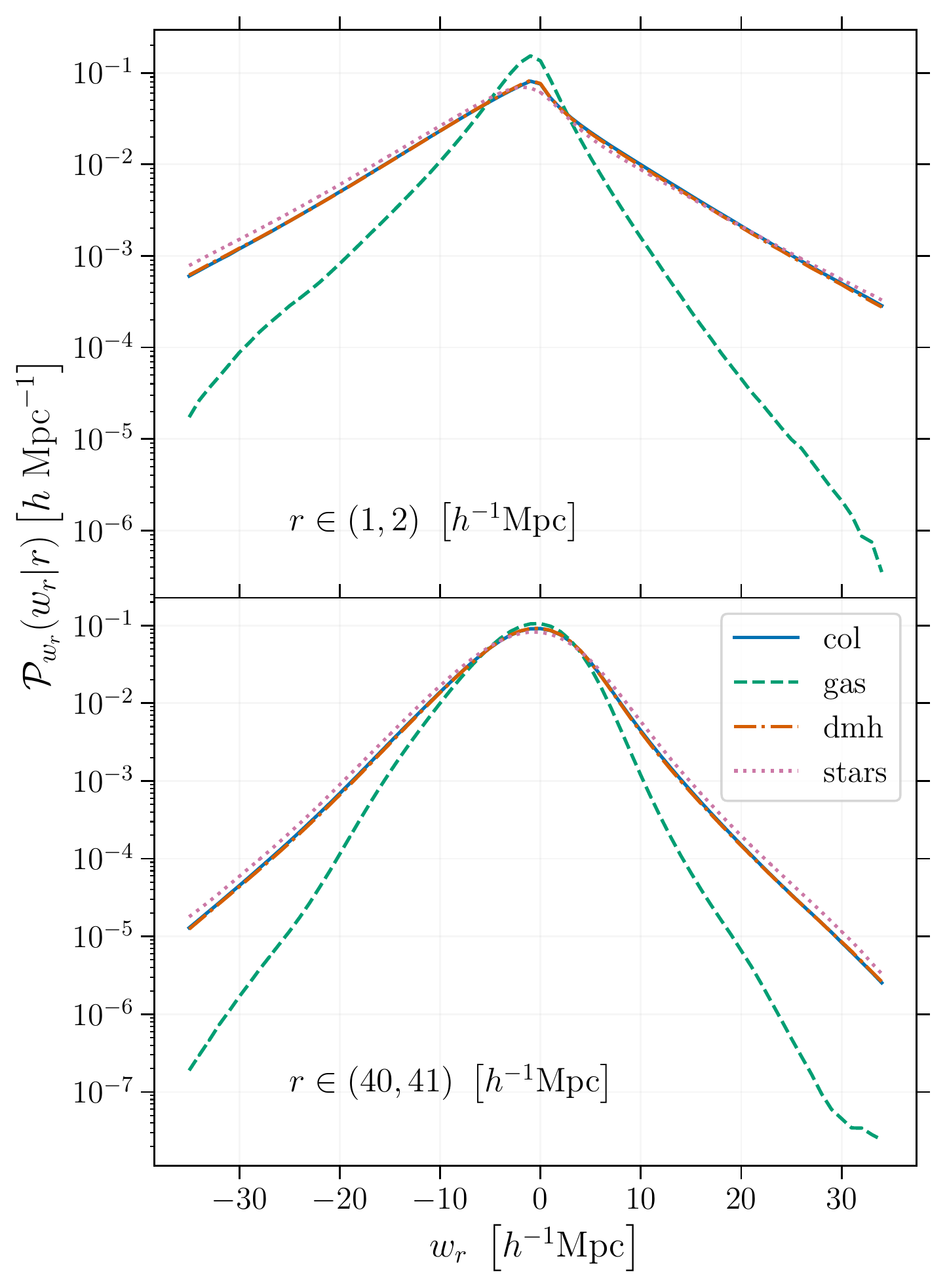}
  \caption{The radial pairwise velocity for different components at $z=0$ from the BAHAMAS simulation with massless neutrinos.  Col (solid) refers to the matter component from the collisionless simulation. Dmh (dash-dotted), gas (dashed) and star (dotted) refers to the dark matter, gas and stellar components from the hydrodynamical simulation respectively. The pair separation of the PDFs are labelled in the bottom left of each panel.}
  \label{fig:rad_pdf}
\end{figure}

The observed galaxy velocities provide a biased view of the unbiased (and unobserved) total matter velocity field, $\bm{v}_{\mathrm{m}}$. This unbiased velocity field can be defined as the fractional sum of its basic components
\begin{equation}
    \bm{v}_{\mathrm{m}} = f_{\mathrm{cdm}}\bm{v}_{\mathrm{cdm}} + f_{\mathrm{b}}\bm{v}_{\mathrm{b}} + f_\nu \bm{v}_{\nu}  \, , 
    \label{eq:matter_vel}
\end{equation}
where $f_{\mathrm{cdm}}\equiv \Omega_{\mathrm{cdm}}/(\Omega_{\mathrm{cdm}} + \Omega_{\mathrm{b}} + \Omega_{\nu})$, $f_{\mathrm{b}}\equiv \Omega_{\mathrm{b}}/(\Omega_{\mathrm{cdm}} + \Omega_{\mathrm{b}} + \Omega_{\nu})$ and $f_\nu\equiv \Omega_{\nu}/(\Omega_{\mathrm{cdm}} + \Omega_{\mathrm{b}} + \Omega_{\nu})$ are the cold dark matter, the baryon and the neutrino fraction respectively.  The velocity of the cold dark matter and the neutrino are denoted by $\bm{v}_{\mathrm{cdm}}$ and $\bm{v}_{\nu}$ respectively. Whereas the velocity of the baryons, $\bm{v}_{\mathrm{b}}$, is further obtained as a fractional sum of the velocities of gas, stars and black holes (BH) 
\begin{equation}
    \bm{v}_{\mathrm{b}} = f_{\mathrm{gas}}\bm{v}_{\mathrm{gas}} + f_{\mathrm{stars}}\bm{v}_{\mathrm{stars}} + f_{\mathrm{BH}}\bm{v}_{\mathrm{BH}}
\end{equation}
where $f_i$ represents the fraction of gas, stars and BH. 
The radial component of the pairwise velocity is
\begin{equation}
    w_{r} = (\bm{v}_2 - \bm{v}_1) \cdot \hat{\bm{r}} \, .
\end{equation}
This can be measured directly from the simulations. In order to build the radial pairwise velocity distribution function (RPVDF) from the simulations, we randomly sample $192^3$ tracer particles. 
In Fig.~\ref{fig:rad_pdf}, we plot the RPVDF of the various components from the BAHAMAS simulation for pairs with a separation of $(1,2)\ h^{-1}\mathrm{Mpc}$ and $(40,41)\ h^{-1}\mathrm{Mpc}$ on top and bottom panels respectively. The solid lines denote the PDF for the matter from the corresponding collisionless simulation. The dashed, dotted and dash-dotted lines are for the gas, stars and dark matter species from the hydrodynamical simulation. It is evident from the PDF that the pairwise velocity information, i.e. all the moments, derived from the gas and DM particles are different at the scales shown. This is important as many studies of the kSZ effect normally assume that the gas perfectly traces the dark matter on large scales.  
It should be noted that $w_r<0$ denotes the pairs which are infalling towards each other, while $w_r>0$ implies that they are moving away from each other. RPVDF of both components are visibly skewed left, while the tails are much heavier for the dark matter component when compared to the gas component.

\section{First moment of radial pairwise velocity}
\label{sec:mean-radial}
\begin{figure}
    \includegraphics[width=0.99\columnwidth]{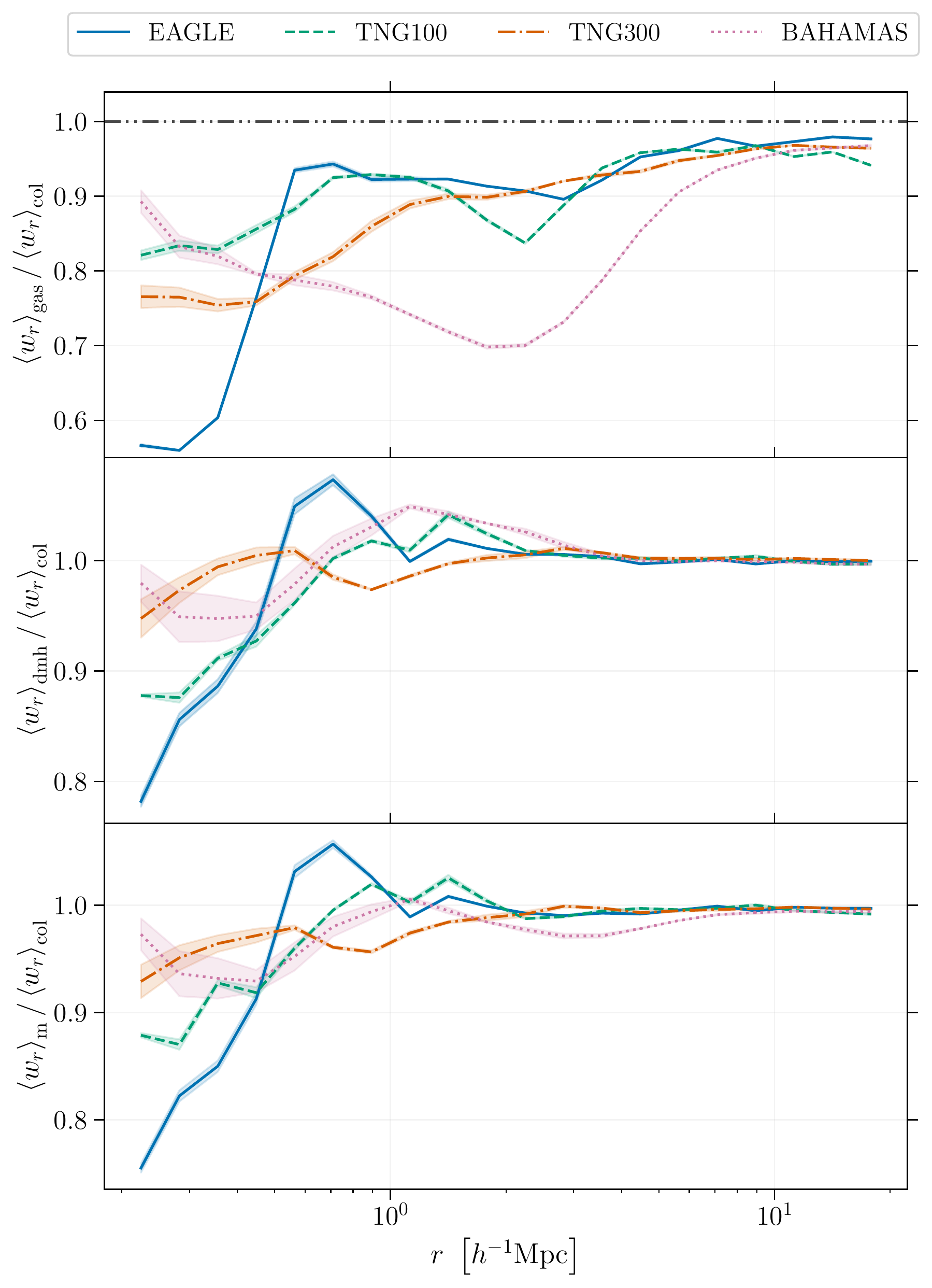} 
  \caption{The baryonic effect on the mean radial pairwise velocity as function of pair separation, at $z=0$ for various hydrodynamical simulations. The top panel shows the ratio between the mean radial pairwise velocity of gas with that of the matter from collisionless simulation, while the middle panel shows the ratio between the mean radial pairwise velocity of the dark matter from full physics run to that of the collisionless simulation. Finally in the bottom panel shows the ratio between the matter from hydrodynamical and collisionless simulations.}
  \label{fig:ratio_z0}
\end{figure}

In this section, we compute the first moment of the radial pairwise velocity from the simulations. In the single stream regime, the mean radial velocity can be defined as: 
\begin{align}
\langle \bm{w}(\bm{r}) \rangle  =\displaystyle \frac{\langle(1+\delta_{1})(1+\delta_{2})(\bm{v}_{2}-\bm{v}_{1})\rangle}{\langle(1+\delta_{1})(1+\delta_{2})\rangle}\;, 
\end{align}
where $\delta$ represents the mass density contrast and $\bm{r}$ gives the pair separation vector. Using standard perturbation theory at leading order, it can be shown that \citep{Fisher95}
\begin{equation}
\langle \bm{w}(\bm{r}) \rangle \simeq \displaystyle - \frac{f}{\pi^2} \, \hat{\boldsymbol{r}} \int_0^{\infty}\!\!\!\!  k \,j_1(k\,r)\,P(k)\, \mathrm{d} k= \langle w_r(r) \rangle \,\hat{\boldsymbol{r}}\, ,
 \label{eq:mean-radial-velocity}
\end{equation}
where $f$ is the growth rate, $j_1(x) = \sin (x)/x^2- \cos (x)/x$ and $P(k)$ is the linear matter power spectrum. The particles in a pair tend to approach each other on average, i.e. $w_r(r)<0$, due to gravitational attraction. In Fig.~\ref{fig:ratio_z0}, we explore the mean radial pairwise velocity for matter, DM and gas components from all the simulations mentioned before. Similar to building the RPVDF, we randomly sample $192^3$ particles for each tracer (except in the case of BH particles) to compute the moments. To quantify the uncertainty of our measurements, we create three such catalogs and use the standard error of the mean.

The top panel shows the effect of different subgrid physics on the gas radial pairwise velocity.  The curves show that all four simulations follow a similar qualitative trend, in that the pairwise velocity of the gas is suppressed relative to the collisionless dark matter, particularly on small scales.  However, the magnitude of this effect varies strongly from simulation to simulation.  At intermediate scales of $1$-$10\ h^{-1}\mathrm{Mpc}$, BAHAMAS (dotted lines) shows the maximal deviation of about 30\% from the assumption that the gas follows the mean velocity of the dark matter.  While both the Illustris-TNG runs show a maximal effect of 10-18\% at the same scales. EAGLE shows an effect of about 10\% (at most) on these intermediate scales.  However on the smallest scales considered, EAGLE shows the largest effect, with the gas pairwise velocity deviating by up to 42\% from the collisionless dark matter.  It should be noted that on all scales considered here the ratio does not go to one, which implies that there is a velocity bias between the dark matter and gas component even on the largest scales that we measure. On below scales of $10\ h^{-1}\mathrm{Mpc}$, the linear velocity bias approximation of mean pairwise velocity clearly breaks down.  
Intriguingly, this holds true for all the simulations we have considered with
varying simulation volumes, thus suggest that this is robust to changes in the box size.

The middle panel of Fig.~\ref{fig:ratio_z0} displays the ratio of the radial pairwise velocity of the dark matter component from the full physics run to the matter component from the collisionless simulation. This comparison tells us how dark matter responds to baryons in the full hydro runs.  On pair separations of about 1-3 $h^{-1}\mathrm{Mpc}$, the BAHAMAS shows a clear back-reaction effect whereby the dark matter component in the full physics run is infalling towards each other at a greater pace than its counterpart in the collisionless simulation. This trend is also seen in the Illustris-TNG100 simulation. 

The bottom panel shows the effect of baryons on the total matter pairwise velocity. The Illustris-TNG runs are within 1\% at scales above 2 $h^{-1}\mathrm{Mpc}$. The matter mean radial pairwise velocity in EAGLE simulation behaves similar to Illustris-TNG at those scales and is affected by $\approx 1\%$ at most. This is also in tandem with the findings in \cite{Hellwing+16}, who find the effect of baryons on redshift-space clustering to be minimal in EAGLE.  However, at small pair separations ($\leq 1\ h^{-1}\mathrm{Mpc}$), matter seems to infall towards each other faster around 0.5-1 $h^{-1}\mathrm{Mpc}$ and this trend reverses quickly at smaller scales. At the intermediate scales, BAHAMAS deviates at about the 2-3\% level. This hints towards the possibility that the redshift-space clustering in BAHAMAS will be affected by baryonic effects to a larger degree than in EAGLE (as shown in Kwan et al., in prep). The fact that BAHAMAS produces a larger effect relative to EAGLE and Illustris-TNG is perhaps not that surprising, as the AGN is more effective at removing baryons from galaxy groups and clusters in BAHAMAS.  This is a result of explicit calibration of the AGN feedback to reproduce the observed baryon fractions of massive systems, whereas neither EAGLE nor Illustris-TNG were calibrated on these data and predict baryon fractions in excess of that observed on mass scales of $\sim10^{14}$ M$_\odot$. By considering the matter pairwise velocity, we study the unbiased velocity field. To directly translate these effects to RSD measurements from redshift surveys, we would need to study the galaxy pairwise velocity statistics which we do not consider in this work.

\begin{figure}
  \centering
  \includegraphics[width=0.99\columnwidth]{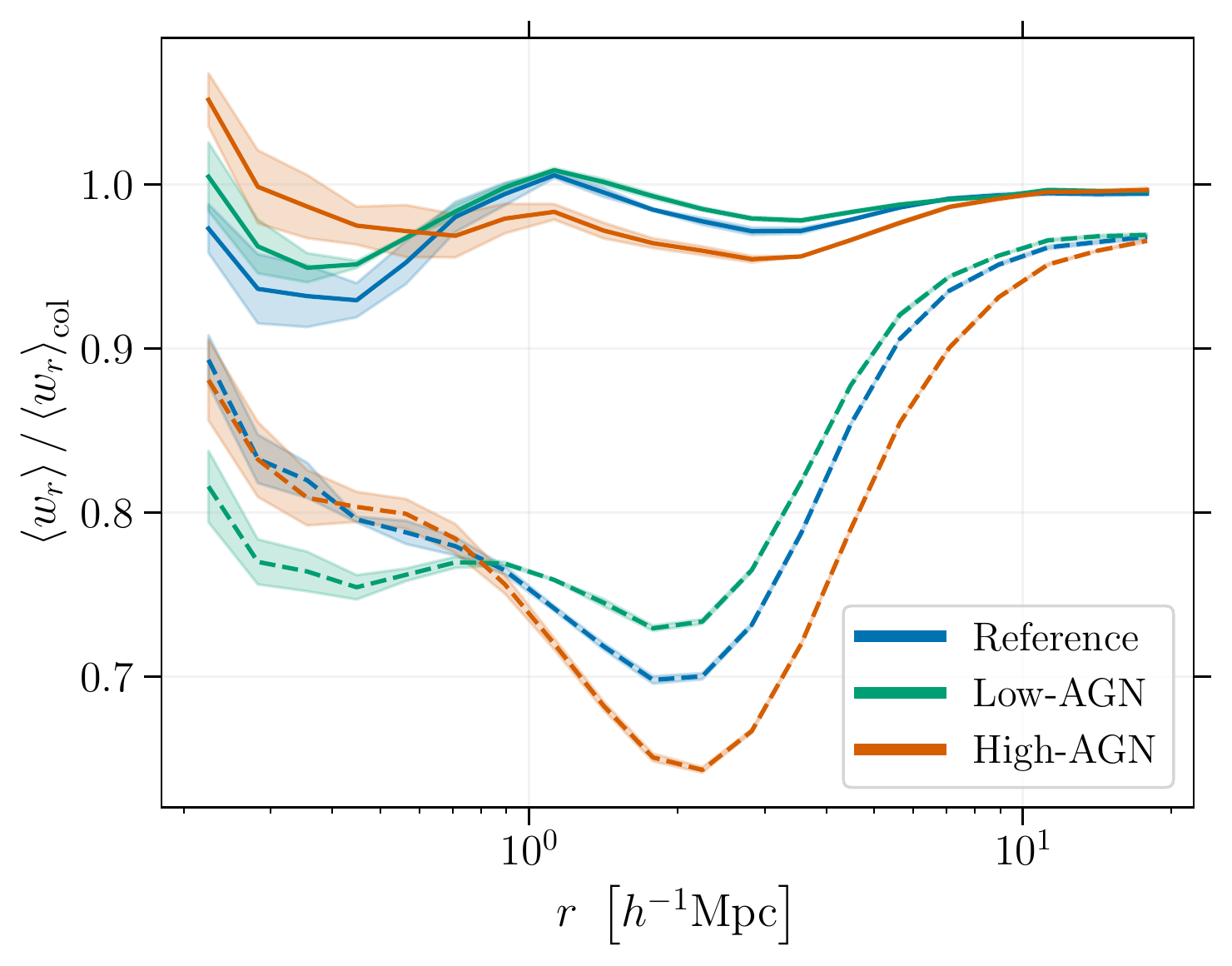}
  \caption{The ratio between the mean pairwise velocity of matter (solid lines) and gas (dashed lines) component with respect to collisionless matter only at $z=0$ for various feedback models in the BAHAMAS suite of simulations.}
  \label{fig:ratio_gas_hi_lo_AGN}
\end{figure}

So far we have seen how the different baryonic models in the simulations affect the velocity statistics.  However, we want to isolate the effect of different physical processes, such as AGN feedback. 
For this purpose, we use two different feedback runs from BAHAMAS with varying AGN subgrid heating temperatures, in addition to the reference run. The `high-AGN' run has $\Delta T_{\mathrm{AGN}} = 10^{8.0}\ \mathrm{K}$, while the `low-AGN' run was run with $\Delta T_{\mathrm{AGN}} = 10^{7.6}\ \mathrm{K}$. These values were chosen so that the simulations roughly bracket the upper and lower bounds of the observed hot gas fraction--halo mass relation inferred from X-ray observations \citep{McCarthy+18}.  They therefore represent a kind of  estimate of the allowed range of behaviours for models with AGN feedback.
In Fig.~\ref{fig:ratio_gas_hi_lo_AGN}, the solid and the dashed lines represent the ratio of matter and gas pairwise velocity with respect to the collisionless matter counterpart, respectively. The gas elements are pushed away from each other more strongly as the AGN temperature increases.  This causes a stronger decrease in the gas radial pairwise velocity for the high-AGN model as can be seen.  Similarly, the matter is also affected in the same manner. The high-AGN feedback causes the matter from the hydrodynamical simulation to deviate further from its counterpart in the collisionless simulation. It should, however, be noted that despite the fact that EAGLE has a higher AGN temperature, the effect of AGN heating is more prominent in BAHAMAS than in EAGLE.  This can be attributed to differences in the mass resolutions of the two simulations, whereby each heating event in BAHAMAS deposits significantly more energy and thus results in a stronger expulsion.  This has also been seen in the case of galaxy clustering information \citep{Foreman+19}.  It would be interesting to run a high-resolution simulation such as EAGLE but to heat a similar volume/mass as that in BAHAMAS to see whether the effects are similar when the feedback is forced to operate in a similar way.

\begin{figure}
  \centering
  \includegraphics[width=0.99\columnwidth]{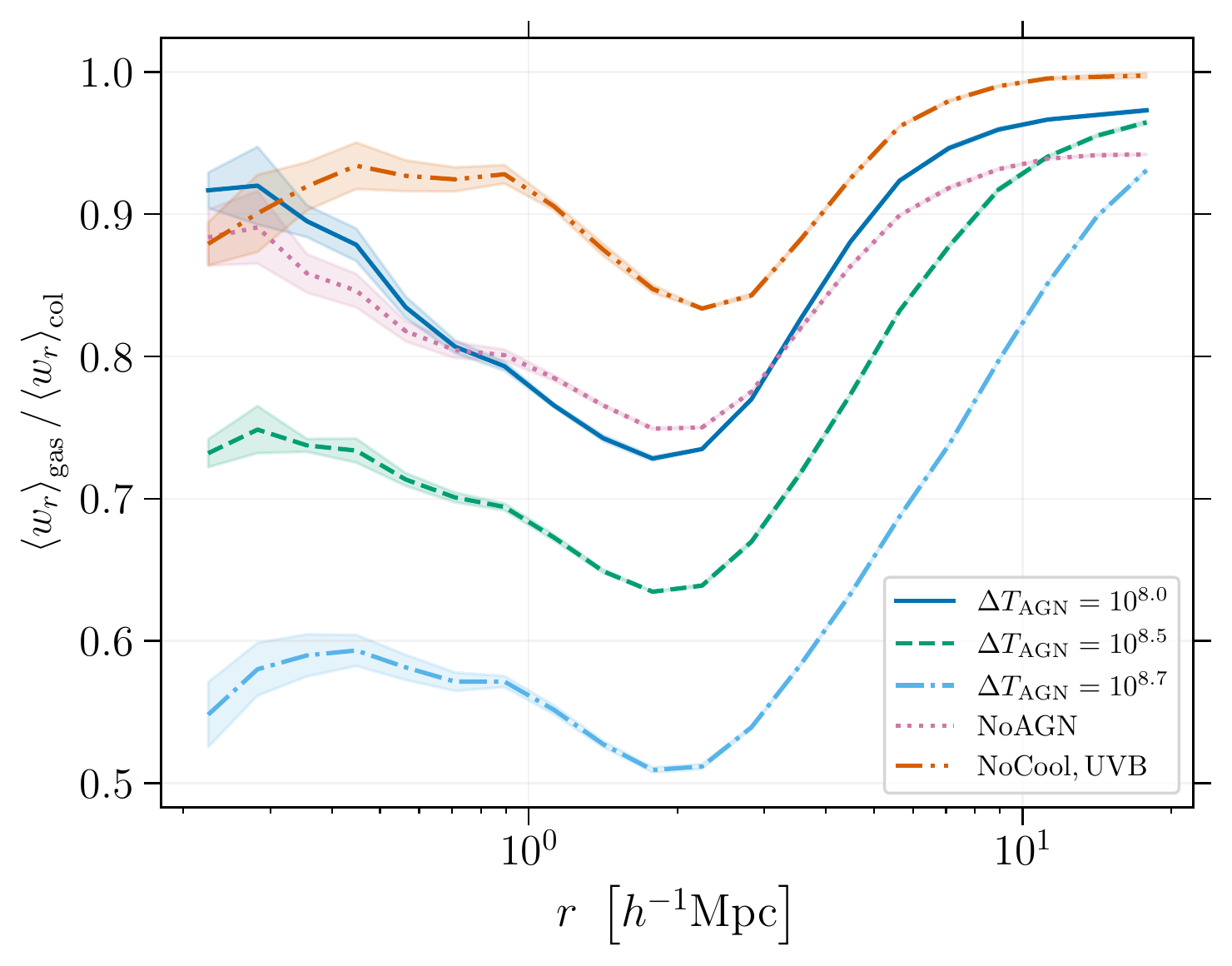}
  \caption{The ratio between the mean pairwise velocity of gas component with respect to collisionless matter at $z=0$ for various runs in the cosmo-OWLS suite of simulations. 
  Dotted line refers to the results from the 'NoAGN' simulation which had the AGN feedback switched off. While dash double dotted line ('NoCool,UVB') refers to the simulation which had no radiative cooling, star formation, or AGN feedback. There is, however, a net photoheating due to a UV background.}
  \label{fig:ratio_gas_CosmoOWLS}
\end{figure}

To further explore the impact of different physical processes, we also make use of the cosmo-OWLS simulations in Fig.~\ref{fig:ratio_gas_CosmoOWLS}. The dashed double-dotted line refers to the `NoCool' simulation in cosmo-OWLS where there is no radiative cooling, star formation, stellar feedback or AGN feedback (there is only net photoheating from a UV/X-ray background).  We see that in this case, the bias is nearly one on scales larger than 10 $h^{-1}\mathrm{Mpc}$.  This implies that it is indeed the physics of galaxy formation that is responsible for the velocity bias on large scales in the previously explored simulations.
Turning on the cooling, star formation and stellar feedback, while keeping AGN feedback turned off (`NoAGN'), we see that this introduces a bias even at scales larger than 10 $h^{-1}\mathrm{Mpc}$. This shows that physical processes like stellar feedback,  prevents the gas from infalling.  The fact that the bias on large scales is similar to that of runs that also include AGN feedback strongly suggests that it is stellar feedback, rather than AGN feedback, that is mainly responsible for the large-scale bias. 

The dash-dotted line shows the effect of AGN feedback in cosmo-OWLS, two models of which have a much higher heating temperature than considered in the case of the BAHAMAS simulation.  As a result, these models clearly expel gas away from each other to a much larger degree.  We note, however, that the two highest heating temperature runs from cosmo-OWLS yield gas fractions significantly lower than observed on the scale of groups and clusters, implying that the feedback is somewhat too aggressive in those runs.

\subsection{Redshift evolution}

\begin{figure}
	\centering
	\includegraphics[width=0.99\columnwidth]{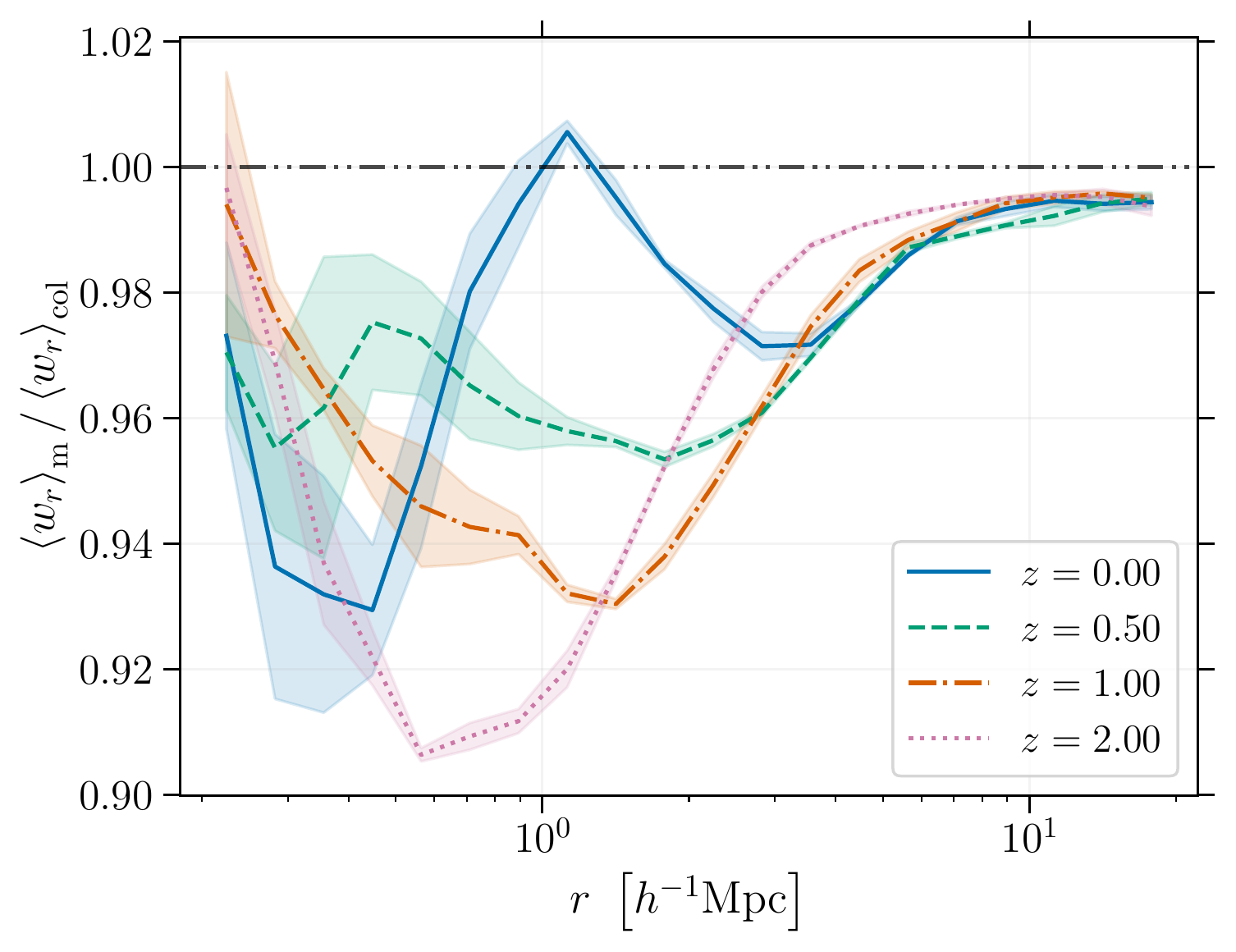}
	\caption{The ratio between the mean pairwise velocity of matter and collisionless matter at various redshifts in the BAHAMAS reference simulation.}
	\label{fig:ratio_redshift_matter}
\end{figure}

We also explore the effect of redshift evolution on the pairwise statistics. For this exercise we use the reference simulation from BAHAMAS with massless neutrinos and measure the mean pairwise velocities at redshifts 0.0, 0.5, 1.0 and 2.0. In Fig.~\ref{fig:ratio_redshift_matter}, we show the effect of baryonic physics on matter mean radial pairwise velocity. The feedback is most efficient at higher redshifts at smaller scales ($r<1\ h^{-1}\mathrm{Mpc}$), reaching a deviation of up to $9\%$ for the matter fluid when compared to its collisionless matter counterpart. At $z=0$ (denoted by solid line), we see the back reaction of DM having an effect on the matter. At scales above $1\ h^{-1}\mathrm{Mpc}$, the ratio reaches a maximal deviation of $\sim 3\%$.  
Thus, these baryonic effects will be important to understand if we are to push the modelling of mean pairwise velocity to non-linear scales and earlier times for forthcoming redshift surveys like Euclid.

\begin{figure}
	\centering
	\includegraphics[width=0.99\columnwidth]{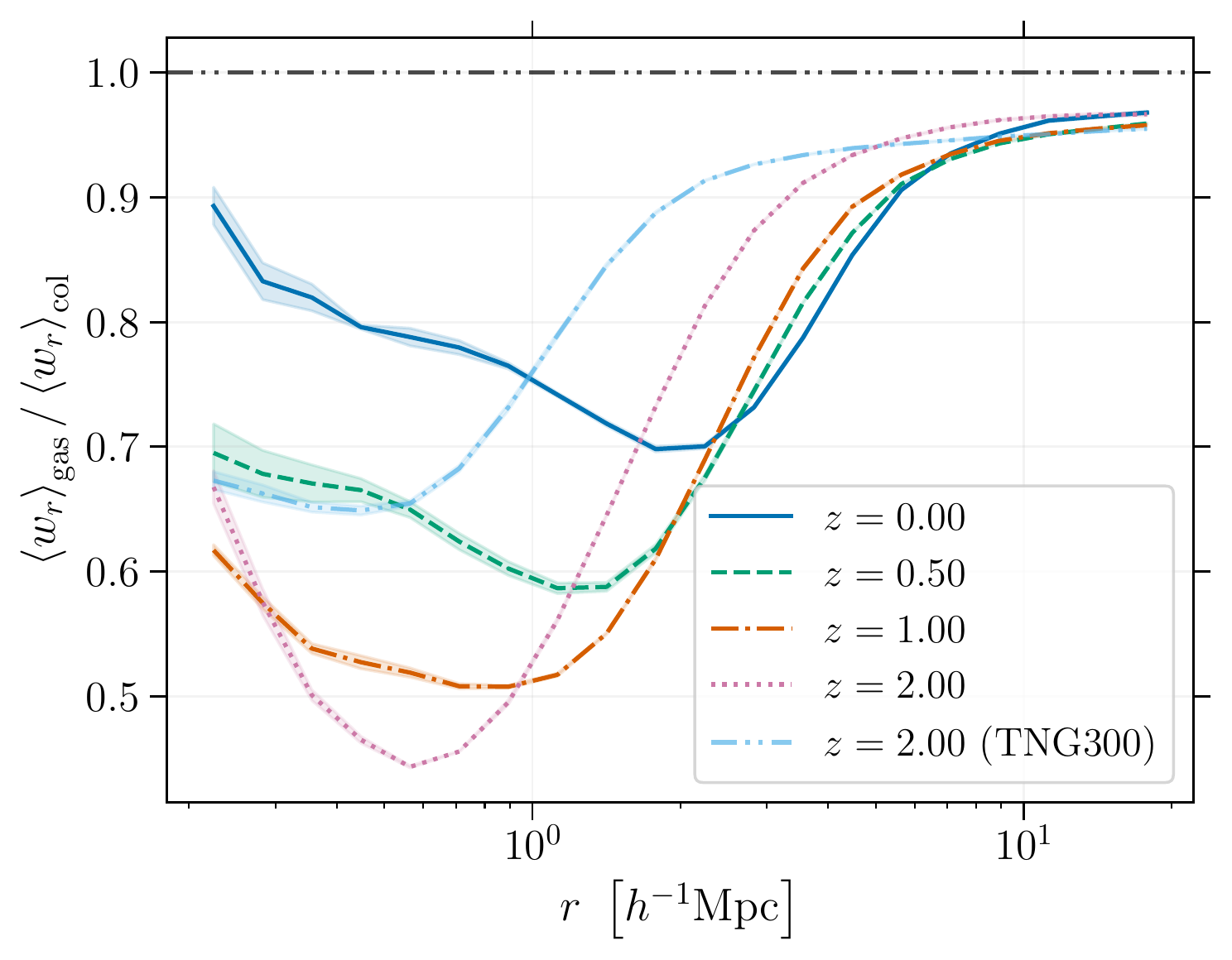}
	\caption{The ratio between the mean pairwise velocity of gas component and collisionless matter at various redshifts in the BAHAMAS reference simulation.}
	\label{fig:ratio_redshift_gas}
\end{figure}

The gas elements show an even more pronounced effect when compared to the matter from the collisionless simulation, as shown in Fig.~\ref{fig:ratio_redshift_gas}. At the highest redshift considered here ($z=2$), the baryonic effects on the gas elements, denoted by dotted lines, strongly affect scales below 3 $h^{-1}\mathrm{Mpc}$.  Moving towards lower redshift, this effect is reduced in amplitude but more extended in scale, being seen on scales as large as $\sim 10\ h^{-1}\mathrm{Mpc}$.  It is again worth highlighting the fact that the velocity bias between the gas and the collisionless matter is below one at all scales considered here and at all times.  This is a clear indication that one needs to be careful about the assumption that mean radial velocity of gas follows that of the dark matter at scales about $20\ h^{-1}\mathrm{Mpc}$ and below, especially for precise measurements in the future.  

For comparison, the dashed double dotted line denotes the trend at $z=2$ from the Illustris-TNG300 simulation.  The effect of AGN feedback in this simulation is strongly reduced compared to BAHAMAS, as was also deduced from the $z=0$ comparison previously.

\subsection{The effects of massive neutrinos}
\label{sec:neutrino-mean}

\begin{figure}
  \centering
  \includegraphics[width=0.99\columnwidth]{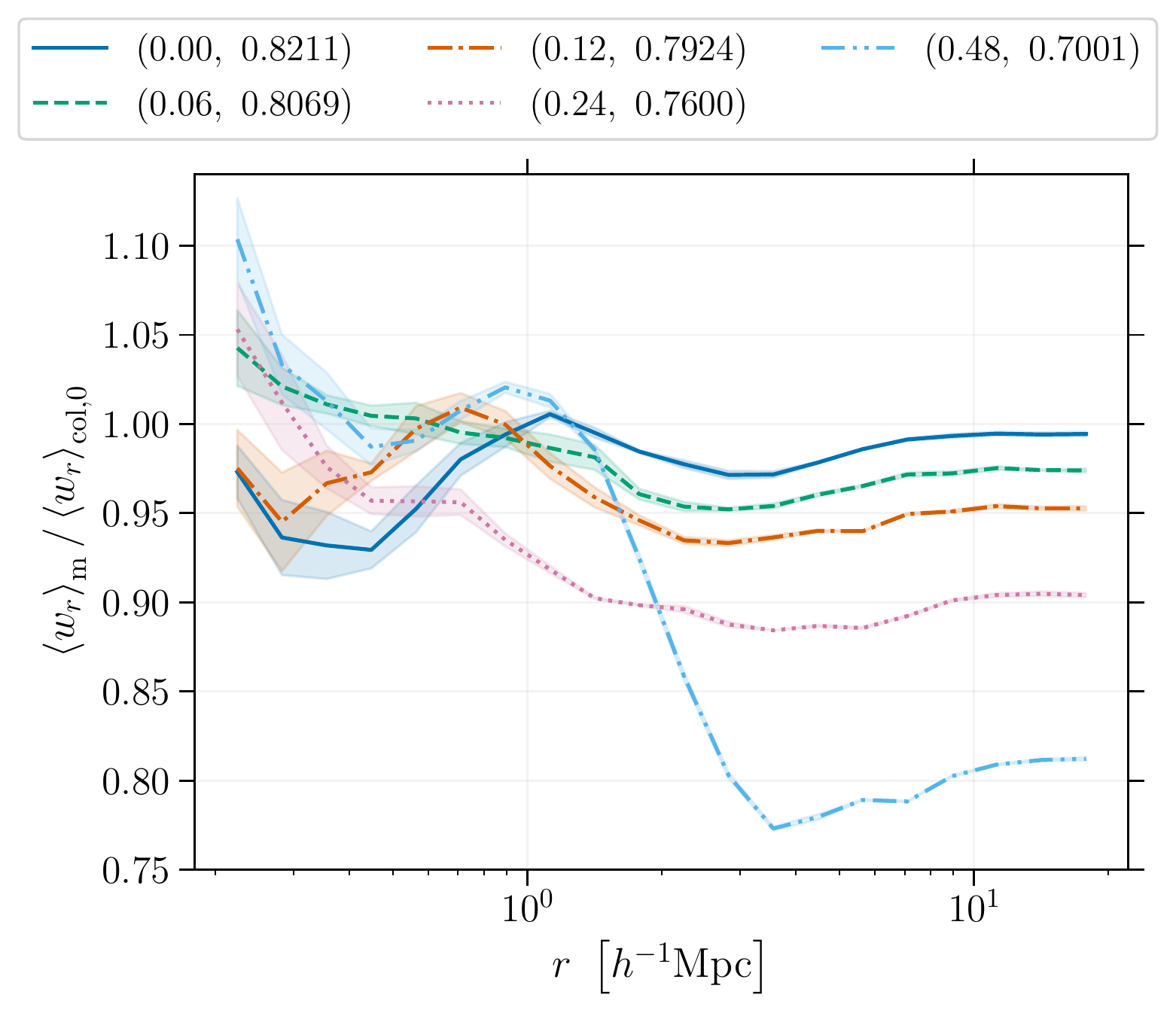}
  \caption{The ratio between the mean pairwise velocity of matter (for various sum of neutrino mass) and for a massless collisionless matter only at $z=0$ in the BAHAMAS suite of simulations. The line styles correspond to different neutrino mass simulations with varying ($M_\nu, \sigma_8$), and $M_\nu$ is given in units of eV.}
  \label{fig:ratio_matter_neutrino}
\end{figure}

Constraining neutrino mass is one of the primary objectives of forthcoming  galaxy and CMB surveys.  One of the main effects of neutrinos on the two-point clustering statistics in Fourier space (i.e. the power spectrum), is the damping of power on scales smaller than the free-streaming scale.  Neutrinos will also affect velocity statistics \citep[e.g.][]{Mueller+15b}.  We focus on the mean radial velocity to exhibit the effects of neutrinos.  Specifically, we show how they affect the matter mean pairwise velocity in Fig.~\ref{fig:ratio_matter_neutrino}. We see that the main effect of neutrinos is to reduce the mean pairwise velocity when compared with a massless neutrino simulation, implying that as the sum of neutrino mass increases, the infall of matter towards each other decreases.  Physically, this is due to  the fact that the neutrino component does not significantly cluster on scales below the free-streaming scale  
which in turn slows the collapse of the dark matter and baryons. 

Considering pair separation scales above 3 $h^{-1}\mathrm{Mpc}$, we can see that the effect reaches approximately 20\% on the matter component for the $M_\nu = 0.48$ eV. This will have important consequences for the RSD signal and hence on the redshift space clustering.  Even for the most stringent of current constraints on the neutrino mass ($M_\nu < 0.12$ eV), the radial pairwise velocity of matter will be affected at the 3-5\% level. This is also encouraging for the future peculiar velocity surveys, which might be able to provide independent constraints on the sum of neutrino masses. 

\begin{figure}
  \centering
  \includegraphics[width=0.99\columnwidth]{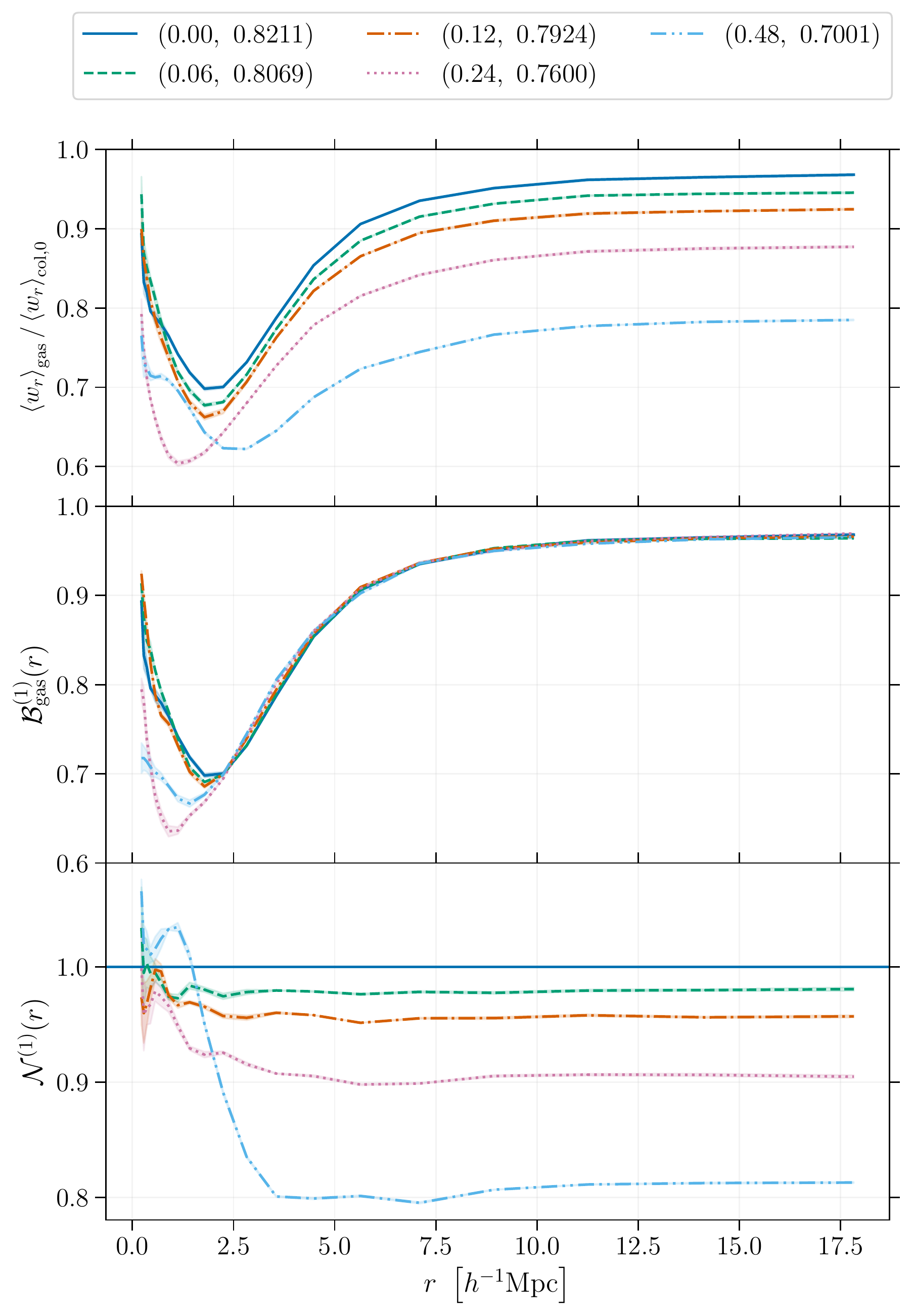}
  \caption{The top panel shows the ratio of the mean radial pairwise velocity between the gas component from BAHAMAS full physics simulation including massive neutrinos and the matter from the BAHAMAS collisionless simulation with a zero mass neutrino. The different line styles represents different neutrino mass simulations. The middle panel shows the effect from purely baryonic effects, while the bottom panel shows the effect of neutrinos  on the mean radial pairwise velocity. The line styles are the same as in Fig.~\ref{fig:ratio_matter_neutrino}.}
  \label{fig:response_functions_gas_neutrino}
\end{figure}

However to decouple the effects of baryons and neutrinos using a single simulation Fig.~\ref{fig:ratio_matter_neutrino} is non-trivial at small scales as the effects are intertwined with each other. Since we have a series of massive neutrino simulations from BAHAMAS (both hydro and collisionless for each neutrino mass), it is possible to disentangle the effects of baryonic physics and massive neutrinos. For this, we introduce the ratio statistics as follows.
\begin{align}
    \frac{\langle w_r \rangle_{i, M_{\nu}}}{\langle w_r \rangle_{\mathrm{col}, 0}} &= \left(\frac{\langle w_r \rangle_{i, M_{\nu}}}{\langle w_r \rangle_{\mathrm{col}, M_{\nu}}}\right) \left(\frac{\langle w_r \rangle_{\mathrm{col}, M_{\nu}}}{\langle w_r \rangle_{\mathrm{col}, 0}}\right)\\
    & = \mathcal{B}^{(1)}_i(r)\ \mathcal{N}^{(1)}(r) \, ,
    \label{eq:response_functions}
\end{align}
where the velocity biases $\mathcal{B}^{(1)}_i(r)$ and $\mathcal{N}^{(1)}(r)$  capture the effects of baryons and neutrinos, respectively, and the subscript $i$ represents either the gas or the matter component. In Fig.~\ref{fig:response_functions_gas_neutrino}, we show the velocity biases due to baryonic effect and massive neutrinos for the gas component, in the middle panel and the  bottom panel respectively. 
The advantage of this approach is that we can treat these effects separately. In the future, one can build emulators for $\mathcal{B}^{(1)}_i(r)$ and $\mathcal{N}^{(1)}(r)$  separately and combine them. The top panel shows the LHS of the equation~(\ref{eq:response_functions}).  Similar to the matter,  massive neutrinos reduce the mean pairwise velocity of gas component at scales above $3\ h^{-1}\mathrm{Mpc}$ and hence the gas velocity bias also decreases as the neutrino mass increases. The function $\mathcal{B}^{(1)}_{\mathrm{gas}}(r)$ is roughly constant above $10\ h^{-1}\mathrm{Mpc}$, below which the baryonic physics starts to have an effect. It can also be seen that the effect from the baryonic processes remain largely unchanged for different neutrino mass cosmology. 
We have also numerically verified that equation~(\ref{eq:response_functions}), holds true for the gas component. At the smallest pair separation considered the relative difference  between the LHS and RHS is 0.1\%, and at the largest separation it is roughly $10^{-5}$\%.

\section{Second moment of radial pairwise velocity}
\label{sec:dispersion-radial}
In this section, we focus on the second moment of the pairwise velocity and check how it is affected by massive neutrinos and the effects of baryons. In the single stream regime, we can define the second moment of the pairwise velocity as 
\begin{align}
\langle \bm{w}(\bm{r})\,\bm{w}(\bm{r}) \rangle =& \displaystyle \frac{\langle(1+\delta_{1})(1+\delta_{2}) (\bm{v}_{2}-\bm{v}_{1}) (\bm{v}_{2}-\bm{v}_{1})\rangle}{\langle (1+\delta_{1})(1+\delta_{2})\rangle} \, .
\end{align}
We are interested in the radial component which can be defined using standard perturbation theory at leading order as 
\begin{align}
\langle (\bm{w} \cdot \hat{\bm{r}})^2 \rangle = 2\left[\sigma^2_v - \psi_{r}(r_{12})\right] \; ,
\end{align}
where  
\begin{equation}
\psi_{r}(r_{12}) = \displaystyle \frac{f^2}{2\pi^2} \int_0^{\infty}  \Biggl[j_0(k\,r_{12}) - 2\, \frac{j_1(k\,r_{12})}{k\,r_{12}}\Biggr]\, P(k)\,\mathrm{d} k\; 
\end{equation}
is the radial velocity correlation function \citep{Gorski88} with $j_0(x) = \sin (x)/x$, and
\begin{equation}
\sigma^2_v = \frac{f^2}{6\pi^2} \int_0^{\infty} \!\!\!\!P(k) \,\mathrm{d} k
\label{eq:sigmav}
\end{equation}
is the one-dimensional velocity dispersion.

\begin{figure}
  \centering
  \includegraphics[width=0.99\columnwidth]{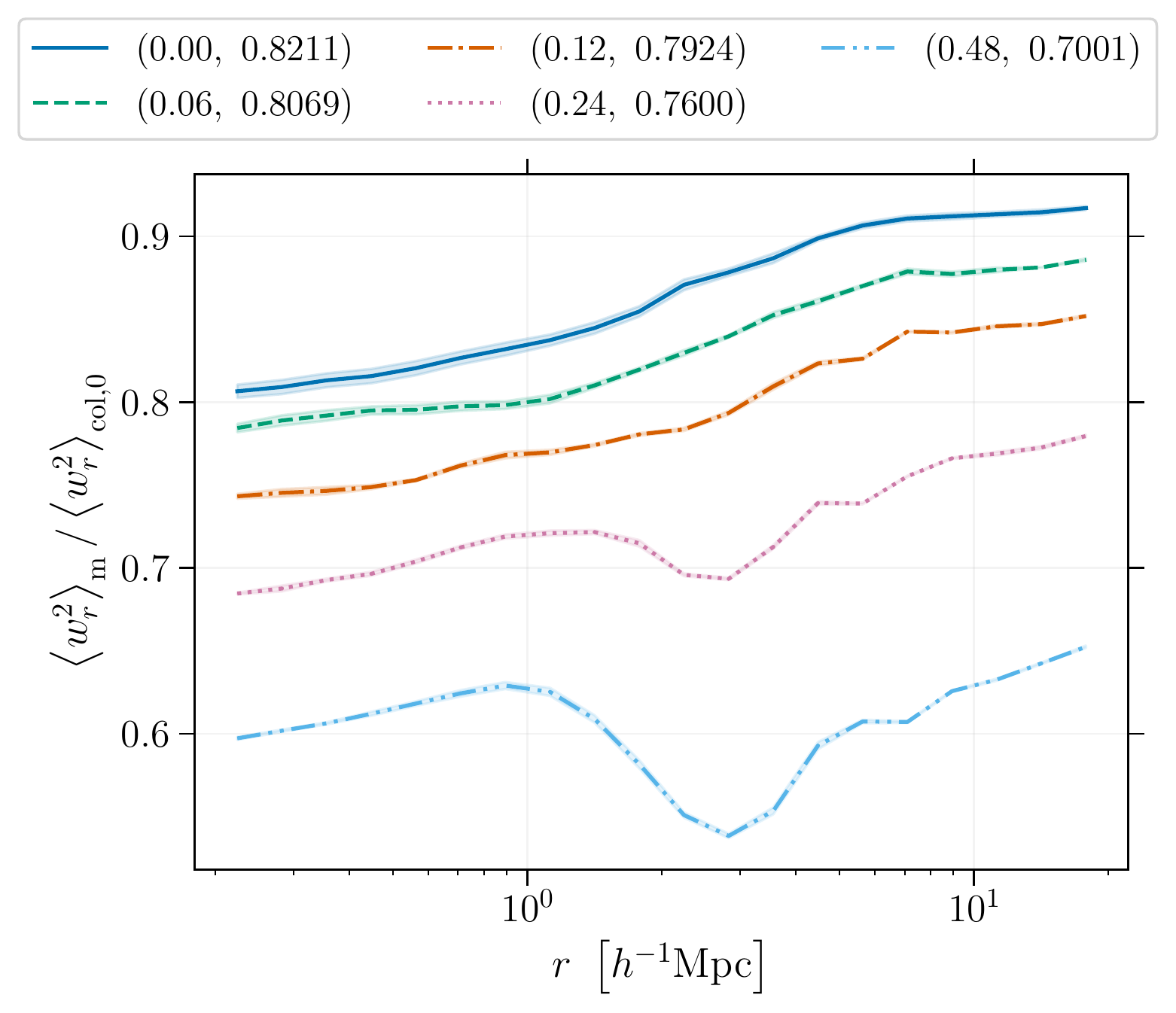}
  \caption{The ratio between the  second moment of the pairwise velocity of matter (for various sum of neutrino mass) from hydrodynamical simulations and the matter from the collisionless simulation with massless neutrinos at $z=0$ in the BAHAMAS suite of simulations. The line styles correspond to variation in ($M_\nu, \sigma_8$), with $M_\nu$ given in units of eV.}
  \label{fig:dispersion_ratio_matter_neutrino}
\end{figure}

In Fig.~\ref{fig:dispersion_ratio_matter_neutrino}, we show the direct measurement of the second moment directly from the BAHAMAS suite of simulations. In the case of the massless neutrinos, the matter in the hydrodynamical simulation has a smaller dispersion compared to the matter in the collisionless simulation. At the largest separation considered here, the second moment is reduced by 8-9\%. As noted already in the first moment, the increasing neutrino mass decreases the velocity dispersion of the radial pairwise velocity. For the most massive neutrino case considered here, the pairwise dispersion is reduced by 25-40\% when compared to the matter from the massless neutrino collisionless simulation. Understanding and accounting for this effect will be important in modelling RSD using the streaming model framework if we want to use clustering analysis at non-linear scales.

\begin{figure}
  \centering
  \includegraphics[width=0.99\columnwidth]{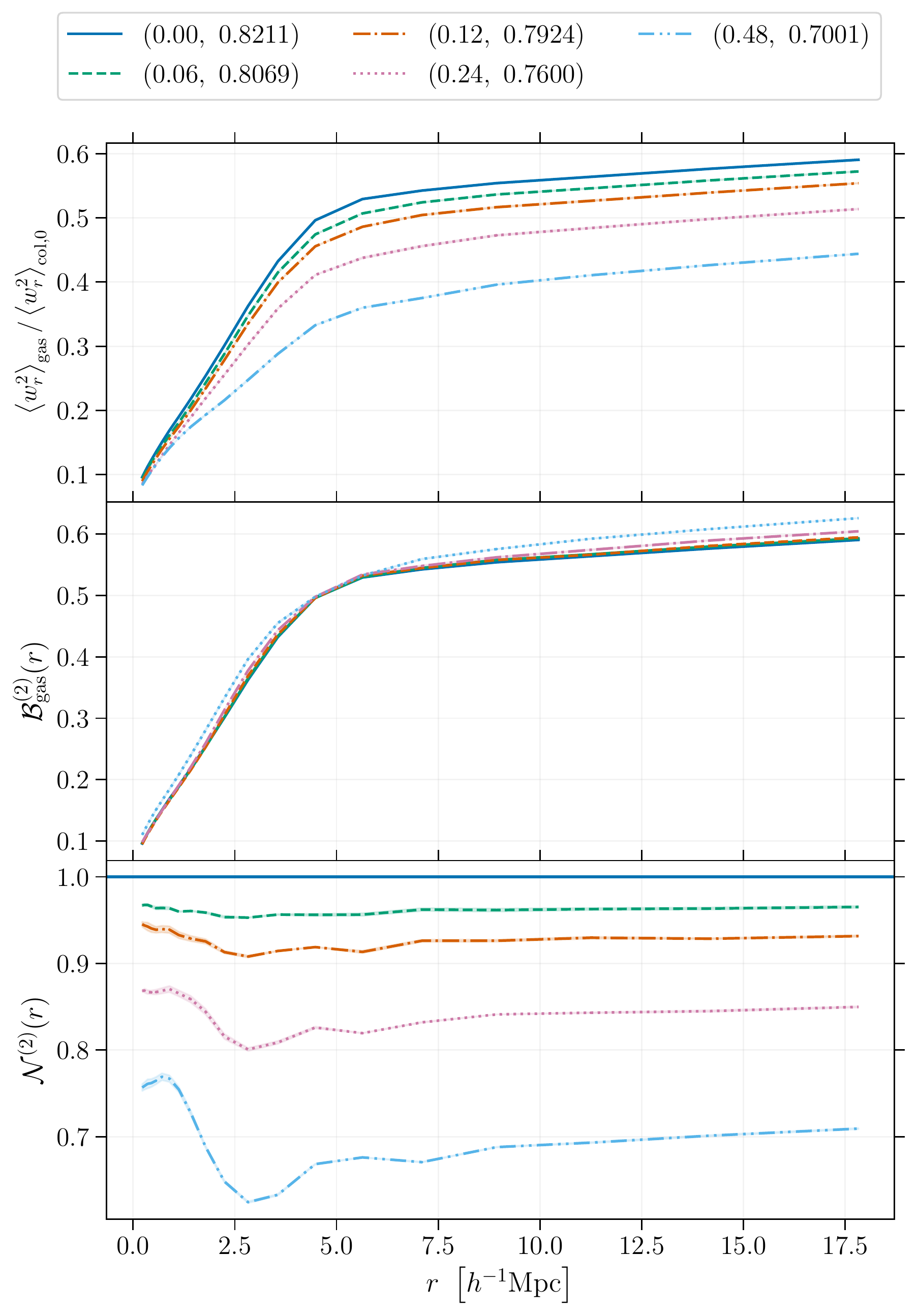}
  \caption{Similar to Fig.~\ref{fig:response_functions_gas_neutrino}, but for the second moment of the radial pairwise velocity.}
  \label{fig:dispersion_response_functions_gas_neutrino}
\end{figure}

To disentangle the effects of baryons and neutrino on the gas dispersion, we can write
\begin{align}
    \frac{\langle w^2_r \rangle_{\mathrm{gas}, M_{\nu}}}{\langle w^2_r \rangle_{\mathrm{col}, 0}} &= \left(\frac{\langle w^2_r \rangle_{\mathrm{gas}, M_{\nu}}}{\langle w^2_r \rangle_{\mathrm{col}, M_{\nu}}}\right) \left(\frac{\langle w^2_r \rangle_{\mathrm{col}, M_{\nu}}}{\langle w^2_r \rangle_{\mathrm{col}, 0}}\right) \nonumber \\
    & = \mathcal{B}^{(2)}_{\mathrm{gas}}(r)\ \mathcal{N}^{(2)}(r) \, ,
    \label{eq:response_functions_gas_dispersion}
\end{align}
where $\mathcal{B}_{\mathrm{gas}}^{(2)}(r)$ and $\mathcal{N}^{(2)}(r)$ are the velocity biases due to baryons and neutrinos, in the context of the pairwise velocity dispersion.
In Fig.~\ref{fig:dispersion_response_functions_gas_neutrino}, we show the LHS (top panel) and RHS (middle and bottom panels) terms of equation~(\ref{eq:response_functions_gas_dispersion}). The top panel shows that the pairwise velocity dispersion of the gas component in the massless neutrino cosmology is significantly less than that of the dark matter, by more than 40\% at all scales considered. The dispersion decreases further as the neutrino mass is increased, as expected. This is encouraging as we can leverage the dispersion measure of the pairwise velocity from kSZ or peculiar velocity measurements to further constrain the summed mass of neutrinos. 

In the middle panel, the effect of baryons is nearly invariant for the different neutrino cosmologies, although for the most massive neutrino case the baryonic effects differ by 1-3\% at pair separations of around 10-20 $h^{-1}\mathrm{Mpc}$ from the massless neutrino case.  The bottom panel effectively shows the impact of neutrinos on the velocity dispersion for dark matter species in the collisionless simulation. The most massive neutrino case causes a decreases of about 30\% even at the largest separations, while a summed neutrino mass of 0.12 eV (dash-dotted line) and 0.24 eV (dotted line) show decreases in the pairwise velocity dispersion of approximately 4\% and 8\%, respectively.

\section{Conclusions}
\label{sec:conclusion}

In this study, we have focused on the imprint of baryons and neutrinos (and their interplay) on the first two moments of the radial pairwise velocity distribution. Understanding these effects will help us to alleviate any potential biases in constraining cosmological parameters, in particular the neutrino mass, from the future surveys.

The assumption that the mean pairwise velocity of gas component follows that of the dark matter is a crucial one undertaken in kSZ analyses. In Fig.~\ref{fig:rad_pdf}, we demonstrated that even on large pair separations, $r \in (40,41)\ h^{-1}\mathrm{Mpc}$, the radial pairwise velocity distribution of the gas component differs from that of the dark matter. 

Focusing on its first moment, 
we demonstrated that different subgrid models lead to different effects on the mean radial pairwise velocity statistics, especially on the very small scales below 1 $h^{-1}\mathrm{Mpc}$, as can been seen in Fig.~\ref{fig:ratio_z0}. We also see that even at pair separations of 15-20 $h^{-1}\mathrm{Mpc}$, there is a pairwise velocity bias between the gas and dark matter. This indicates that the assumption that the mean pairwise velocity of gas follows that of dark matter breaks down at these scales.

We further studied the effect of AGN feedback in particular on the mean pairwise velocity in Fig.~\ref{fig:ratio_gas_hi_lo_AGN} using the BAHAMAS simulations, finding that more energetic AGN heating pushes the matter away leading to a decrease in the mean infall of material.  In Fig.~\ref{fig:ratio_gas_CosmoOWLS}, we studied the effect of different baryonic processes using cosmo-OWLS suite of simulation. The assumption that the gas follows the dark matter (above scales of 10 $h^{-1}\mathrm{Mpc}$) is valid only in the case when all non-gravitational physical processes like radiative cooling, star formation and stellar and AGN feedback are switched off. 
In the cases when those physical processes were switched on, the assumption breaks down for the pair separations we have considered. 
The source of the large-scale velocity bias appears to be driven by the stellar feedback rather the AGN feedback as suggested in Fig.~\ref{fig:ratio_gas_CosmoOWLS}. Turning AGN feedback on does not significantly alter this, but it does greatly affect intermediate scales. Thus the strength of the variation changes according to the subgrid physics considered.

The impact of baryonic process at different redshifts is studied using the BAHAMAS reference simulation in Figs.~\ref{fig:ratio_redshift_matter} and \ref{fig:ratio_redshift_gas}. We see that even at the highest redshift considered in our study $z=2$, the baryonic processes introduce one percent level impact on matter mean pairwise velocity at scales above 10 $h^{-1}\mathrm{Mpc}$. In the case of gas component, the impact is more prominent and introduces 4-5\% change in the mean velocity with respect to the matter in a gravity-only calculation.

We studied the effect of massive neutrinos on mean radial pairwise velocity using the BAHAMAS suite of simulations. We showed that the matter mean pairwise velocity decreases as the summed neutrino mass increases, in Fig.~\ref{fig:ratio_matter_neutrino}.  Though we studied the (unbiased) matter velocity field, these results suggest that the radial pairwise velocity could be used to potentially constrain neutrino mass from peculiar velocity surveys in the future, and in addition these effects could be important in modelling RSD using the streaming model framework in the presence of massive neutrinos. In Fig.~\ref{fig:response_functions_gas_neutrino}, we disentangled the baryonic and massive neutrino effects on the mean radial pairwise velocity of gas component as introduced in equation~(\ref{eq:response_functions}). For the most massive neutrino considered in this work ($M_\nu = 0.48$ eV), we found that the mean radial pairwise velocity of gas decreases by roughly 20\% when compared with dark matter in the massless neutrino simulation. The baryonic effect is nearly invariant when considering different neutrino mass simulations. 
Finally, we demonstrated the effect of neutrinos on the second moment of the radial pairwise velocity for both matter and and gas components, as shown in Figs.~\ref{fig:dispersion_ratio_matter_neutrino} and \ref{fig:dispersion_response_functions_gas_neutrino} respectively. Similar to the mean radial pairwise velocity, the second moment also decreases with increasing neutrino mass. The matter pairwise velocity is reduced by $\sim$15 \% for $M_\nu = 0.12$ eV when compared to the massless neutrino case at pair separations of 10-20 $h^{-1}\mathrm{Mpc}$.  At the same separation, for the highest neutrino mass considered the impact is reduced by $\sim$35 \%. This points towards the possibility of utilising the pairwise dispersion (as a function of pair separation) to constrain neutrino mass from either future peculiar velocity surveys or future CMB surveys using kSZ effect.  The second moment would also be beneficial for breaking degeneracies between cosmological parameters.  For example, the mean pairwise velocity scales as $f\sigma^2_8$, while the second moment scales as $(f\sigma_8)^2$. Direct application of our results to either peculiar velocity surveys or kSZ would require us to study the effect separately on galaxies/haloes, which we reserve for a future study.

Thus, we have seen how different feedback models affect the moments of the pairwise velocity to varying degree. With the forthcoming peculiar velocity and CMB surveys, understanding these systematic effects from baryons and neutrinos will be essential for constraining the cosmological parameters using pairwise velocity accurately and precisely.

\begin{acknowledgements}
We thank the referee for their comments. JK and NA acknowledge funding for the ByoPiC project from the European Research Council (ERC) under the European Union's Horizon 2020 research and innovation program grant agreement ERC-2015-AdG 695561. IM has received funding from the ERC under the European Union’s Horizon 2020 research and innovation programme (grant agreement No 769130).
We are thankful  to  the  community for developing  and  maintaining open-source software packages extensively used in our work, namely \textsc{Cython} \citep{cython}, \textsc{Matplotlib} \citep{matplotlib} and \textsc{Numpy} \citep{numpy}.
\end{acknowledgements}

\setlength{\bibhang}{2.0em}
\setlength\labelwidth{0.0em}
\bibliographystyle{aa}
\bibliography{main}

\end{document}